\documentclass[8pt]{article}
\usepackage{spconf,amsmath,graphicx}
\usepackage[printonlyused]{acronym} 
\usepackage{psfrag}
\usepackage[tight]{subfigure}
\usepackage{color} 

\usepackage{pgfplotstable}
\usepackage{tikz}

\usepackage{pgfplots}
\pgfplotsset{compat=1.3}

\definecolor{LMSdarkgreen}{rgb}{0,0.5,0}

\acrodef{ASR}{Automatic Speech Recognition}
\acrodef{BSE}{Blind Source Extraction}
\acrodef{BSS}{Blind Source Separation}
\acrodef{DoA}{Direction of Arrival}
\acrodef{GCC}{Generalized Cross-Correlation}
\acrodef{MSC}{Magnitude Squared Coherence}
\acrodef{RIR}{Room Impulse Response}
\acrodef{PSD}{Power Spectral Density}
\acrodef{PSDs}{Power Spectral Densities}
\acrodef{SIR}{Signal-to-Interference Ratio}
\acrodef{ULA}{Uniform Linear Array}
\acrodef{VAD}{Voice Activity Detection}

\title{Improving Blind Source Separation Performance By Adaptive Array Geometries For Humanoid Robots}

\makeatletter
\def\name#1{\gdef\@name{#1\\}}
\makeatother
\name{{\em Hendrik Barfuss} and {\em Walter Kellermann}\thanks{The research leading to these results has received funding from the European Union's Seventh Framework Programme (FP7/2007-2013) under grant agreement n$^\mathsf{o}$ 609465.}}
\address{Multimedia Communications and Signal Processing,\\
         University of Erlangen-Nuremberg\\
         Cauerstr. 7, 91058 Erlangen, Germany \\
         {\small \tt \{barfuss,wk\}@LNT.de}}

\begin{document}
%
\maketitle
%
\begin{abstract}
In this paper, the concept of an adaptation algorithm is proposed, 
which can be used to blindly adapt the microphone array geometry of a humanoid robot 
such that the performance of the underlying signal separation algorithm is improved.
As a decisive feature, an online performance measure for blind source separation is introduced which allows a robust and reliable estimation of the instantaneous separation performance 
based on currently observable data.
Experimental results from a simulated environment confirm the efficacy of the concept.
\end{abstract}

\section{Introduction}
\label{sec:introduction}
For natural human/robot  interaction, robot audition should support speech communication even if the human is at a distance of several meters.
As an example, we may consider a robot acting as an information point in public spaces, e.g., a welcoming robot in a hotel lobby.
In such a scenario, the robot will be located at some distance from his desired human communication partner while other interfering speakers and background noise may be active at the same time. 
Therefore, a key problem in robot audition is to extract the desired source signal from the mixture of desired and interfering sources and background noise.
For such scenarios, a two-channel \ac{BSE} approach has been proposed by Reindl et al.~\cite{lnt2012-62}, showing promising results in noisy living-room-like environments such as the one of the  $\mathrm{PASCAL \, CHiME}$ challenge \cite{Barker:2013}.~This \ac{BSE} approach consists of several steps: 
In the first step, the \ac{DoA} of the desired signal needs to be estimated using methods as, e.g., described in \cite{Knapp:1976, Nesta:2009fk, lnt2011-3}.
%
In the second step, the obtained \ac{DoA} is used to estimate the interference and noise components as proposed by Zheng et al.~\cite{Zheng:2009fk}. 
This approach utilizes the $\mathrm{TRINICON}$ ($\mathrm{TRI}$ple-N Independent $\mathrm{CO}$mponent Analysis for $\mathrm{CON}$volutive mixtures) \ac{BSS} algorithm, introduced in \cite{lnt2004-6},
for separating all desired source signal components including correlated echoes from all interference and noise components.
Finally, given the noise estimate from the second step, Wiener-type spectral enhancement filters are applied to the microphone signals in order to suppress all undesired signal components.
Clearly, the key to this algorithm is to obtain a good noise estimate, since the quality of the latter determines the performance of Wiener-type filters and, thus, of the entire extraction algorithm.
As a decisive advantage of this scheme, no source activity detection or estimation nor any source modelling is necessary, and no knowledge of the array geometry is required.

Using a humanoid robot offers the opportunity to place microphones not only on the head, but also on the movable limbs, making it possible to change the aperture size of the microphone array by letting the robot stretch out its arms 
or pull them back in. This movement could be incorporated into a welcoming gesture of the robot. 
Since the beamwidth of a beamformer is directly related to the array length, see. e.g., \cite{Trees:2002fk}, 
this mechanism can be used to increase the separation performance of the \ac{BSS} algorithm if interfering sources are very close to the desired source. 
As a consequence, the performance of the signal extraction algorithm increases and the robot is enabled to focus on a desired source in such a scenario.

\begin{figure}[h]
  \centering
  \psfrag{1}[c][c][.85]{\scriptsize Humanoid robot}
  \psfrag{2}[c][cr][.85]{\parbox{1.5cm}{\centering \scriptsize Desired speaker}}
  \psfrag{3}[c][cl][.85]{\parbox{1.5cm}{\centering \scriptsize Interfering speaker}}
  \includegraphics[width=.35\textwidth]{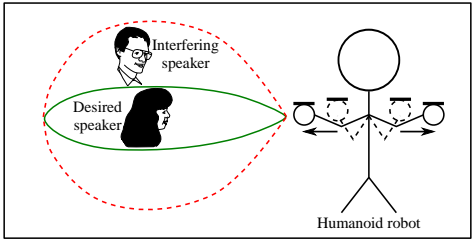}
  \caption{Illustration of the underlying idea of the proposed adaptation algorithm. Microphones mounted on the hands of the robot offer the possibility to increase the array aperture, and thus, the separation performance of the underlying \ac{BSS} algorithm.}
  \label{fig:ilustrationMainPrinciple}
\end{figure}
An illustration of this concept is given in Fig.~\ref{fig:ilustrationMainPrinciple}. Here, the dashed red ellipsoid denotes the case where the robot cannot distinguish between the desired and interfering sources
due to an array aperture which is too small. Increasing the array aperture size enables the robot to focus on the desired source, as denoted by the solid green ellipsoid.

In this paper, an algorithm is proposed which iteratively adapts the microphone distances of a uniform linear array, such that the separation performance of an underlying \ac{BSS} algorithm is optimized. 
As a key ingredient, an online performance measure is introduced, which estimates the separation performance of the \ac{BSS} algorithm blindly and reliably based on currently observable data.
The efficacy of the proposed algorithm is verified by experiments using the $\mathrm{TRINICON}$ \ac{BSS} algorithm in a simulated acoustic environment for a three-sensor linear microphone array.
%
The paper is structured as follows. In Section~\ref{sec:proposedAdaptationAlgorithm}, the adaptation algorithm is introduced. The performance measure and experimental results are presented in Section~\ref{sec:experiments},
showing that the algorithm is capable of adapting the microphone distances of a linear microphone array in a multi-speaker scenario such that the separation performance of the underlying \ac{BSS} algorithm is improved. 
The paper is concluded by a summary of the results and an outlook on future work in Section~\ref{sec:summary_outlook}.

\section{Proposed Adaptation Algorithm}
\label{sec:proposedAdaptationAlgorithm}
\begin{figure}[htp]
  \centering
  \psfrag{d1}[c][c]{$d_1^{(j)}$}
  \psfrag{d2}[c][c]{$d_2^{(j)}$}
  \psfrag{a1}[cr][cr]{sub-array $1$}
  \psfrag{a2}[cl][cl]{sub-array $2$}
  \includegraphics[width=.25\textwidth]{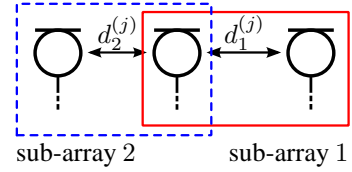}
  \caption{Illustration of the employed sub-arrays for the case of a three-sensor linear array.}
  \label{fig:sub-arrays}
\end{figure}

The main concept of the adaptation algorithm for the array geometry is based on the fact that the microphone array can be configured as two sub-arrays. 
The adaptation of the two microphone spacings $d_i^{(j)}, \, i \in \{1,2\}$ at the $j$-th iteration step of the array geometry adaptation is then based on the separation performance obtained by performing the \ac{BSS} for each of the two sub-arrays. 
In the following, the array geometry adaptation algorithm is illustrated by using a three-sensor linear array which is configured as two two-sensor sub-arrays, as illustrated in Fig.~\ref{fig:sub-arrays}. 
For this illustration, the acoustic scenario is assumed to remain time-invariant. Suggested microphone positions are given by one microphone attached to each of the robot's hands and one microphone located at the robot's torso.
Given this configuration, each sub-array consists of the center microphone which stays at a fixed position and of the right-hand and left-hand microphone, represented by the solid red and dashed blue box in Fig.~\ref{fig:sub-arrays}. 

Without loss of generality, we assume that the initial microphone spacings $d_i^{(0)}, \, i\in \{1,2\}$ are chosen such that $d_1^{(0)} < d_2^{(0)}$.

In the first phase of each geometry adaptation step $j$, the \ac{BSS} algorithm for each microphone pair has to identify the optimum \ac{BSS} filters for each of the current microphone spacings. Therefore, the input signal pairs
are processed block-wise until convergence of the \ac{BSS} algorithm,  see, e.g., the block-online adaptation of $\mathrm{TRINICON}$~\cite{lnt2006-19}.
For the experiments presented in Section~\ref{sec:experiments}, segments of the microphone signals of length of ten seconds were used for each \ac{BSS} adaptation phase.
%
After each \ac{BSS} adaptation phase, a performance measure $f(d_i^{(j)}), \, i \in \{1,2\}$ is computed which characterizes the separation performance of the corresponding $i$-th sub-array.
The employed performance measure is discussed in Section~\ref{sec:experiments} in more detail. Essentially, it estimates the correlation between the two output channels of the \ac{BSS} algorithm.
Thus, small values of $f(d_i^{(j)})$ correspond to a good separation performance.

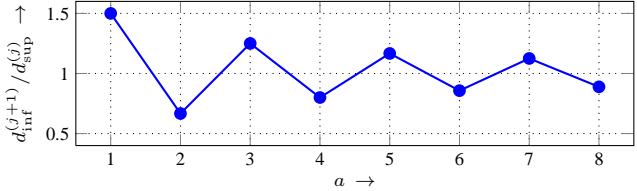
\begin{figure}[h]
  \scriptsize
  \centering
  \begin{tikzpicture}[scale=1]
    \begin{axis}[
      width=9cm,height=3.5cm,grid=major,grid style = {dotted,black},
      ylabel={$d_\mathrm{inf}^{(j+1)}/d_\mathrm{sup}^{(j)} \,\rightarrow$},
      xlabel={$a \,\rightarrow$},
      xtick={1,2,3,4,5,6,7,8},
      xticklabels={1,2,3,4,5,6,7,8},
      ytick={.5,1,1.5},
      yticklabels={0.5,1,1.5},
      ymin=0.4,ymax=1.6,xmin=.5,xmax=8.5]
      \addplot[thick,blue,solid,mark=*] table [x index=0, y index=1]{adaptation_strategy.dat};		
    \end{axis}
  \end{tikzpicture}
  \caption{Illustration of the adaptation of the microphone spacing $d^{(j+1)}_\mathrm{inf}$ of the worse performing sub-array. $d^{(j+1)}_\mathrm{inf}$ converges to $d^{(j)}_\mathrm{sup}$.}
  \label{fig:adaptation_strategy}
 \end{figure}

In the second phase of each geometry adaptation step $j$, based on the obtained performance measures $f(d_i^{(j)})$ of the converged \ac{BSS} algorithms of the two current microphone array geometries, the microphone distances $d_i^{(j)}, \, i \in \{1,2\}$ are adapted as follows.
In general, the microphone spacing of the worse performing sub-array is adapted such that 
\begin{equation}
  d_\mathrm{inf}^{(j+1)}=\left(1+\frac{(-1)^{a+1}}{a+1}\right) d_\mathrm{sup}^{(j)}, \, a \in \{1,2,...,a_\mathrm{max}\},
  \label{eq:adaptationStrategy}
\end{equation}
where $d_\mathrm{inf}$ and $d_\mathrm{sup}$ represent microphone spacings with inferior and superior performance at iteration j, respectively, and $a_\mathrm{max}$ is the value of $a$
where the distance between $d_\mathrm{sup}^{(j)}$ and $d_\mathrm{inf}^{(j+1)}$ is smaller than a threshold $\epsilon$: $|d_\mathrm{sup}^{(j)}-d_\mathrm{inf}^{(j+1)}| \leq \epsilon$.
The idea behind this update strategy in (\ref{eq:adaptationStrategy}) is to always create a new competitor to the currently best performing sub-array. 
In Fig.~\ref{fig:adaptation_strategy}, the evolution of the adapted stepsize for different values of $a$ is illustrated. As can be seen, the adapted values $d_\mathrm{inf}^{(j+1)}$ converge to the currently best performing spacing $d_\mathrm{sup}^{(j)}$ with increasing $a$.

In order to react to a performance degradation of the superior micophone array geometry due to, e.g., the time-variance of the scene, 
the microphone spacing of the worse-performing sub-array is increased to a large aperture to find a better array geometry. 
Thus, in case of a decreasing separation performance of the superior sub-array over a number $t_\mathrm{max}$ of geometry adaptation steps $j$, 
i.e., $f(d^{(j+t)}_\mathrm{sup}) > f(d^{(j)}_\mathrm{sup}), \, t \in \{1,...,t_\mathrm{max}\}$,
the spacing of the worse performing sub-array is doubled: $d^{(j+t_\mathrm{max}+1)}_\mathrm{inf} = 2 d^{(j+t_\mathrm{max})}_\mathrm{inf}$.

\begin{figure*}[htp] 
  \centering
  \scriptsize
  \psfrag{t0}[c][c]{\parbox{3cm}{\centering {\small Initialization:}\\ \vspace{1mm} $d_1^{(0)} < d_2^{(0)}$, $a_{\{1,2\}}=1$}}
  \psfrag{t1}[c][c][.9]{\parbox{3cm}{\small \centering \acs{BSS} adaptation for the new array geometries}}
  \psfrag{t2}[c][c][.9]{\parbox{3 cm}{\centering  $a_2 = 1$,  $d_\mathrm{sup}^{(j)} = d_2^{(j)}$, $f(d_\mathrm{sup}^{(j)}) = f(d_2^{(j)})$ }}
  \psfrag{t3}[c][c][.9]{\parbox{3 cm}{\centering  $a_1 = 1$,  $d_\mathrm{sup}^{(j)} = d_1^{(j)}$, $f(d_\mathrm{sup}^{(j)}) = f(d_1^{(j)})$ }}
  \psfrag{t5}[c][c]{\parbox{3cm}{\centering  $d_2^{(j+1)}=2 d_2^{(j)}$}}
  \psfrag{t7}[c][c]{\parbox{3cm}{\centering  $d_1^{(j+1)}=2 d_1^{(j)}$}}
  \psfrag{t4}[c][c][.8]{\parbox{3cm}{\centering  $d_2^{(j+1)}=2 d_\mathrm{sup}^{(j-t_\mathrm{max})}$,\vspace{2mm} $t=1$}}
  \psfrag{t6}[c][c][.8]{\parbox{3cm}{\centering  $d_1^{(j+1)}=2 d_\mathrm{sup}^{(j-t_\mathrm{max})}$,\vspace{2mm} $t=1$}}
  \psfrag{t4}[c][c][.55]{\parbox{4cm}{\centering  $d_2^{(j+1)}=\left(1+\frac{(-1)^{a_2+1}}{a_2+1}\right) d_1^{(j)}$, \\ \vspace{1mm} $a_2=a_2+1$}}
  \psfrag{t6}[c][c][.55]{\parbox{4cm}{\centering  $d_1^{(j+1)}=\left(1+\frac{(-1)^{a_1+1}}{a_1+1}\right) d_2^{(j)}$, \\ \vspace{1mm} $a_1=a_1+1$}}
  \psfrag{t8}[c][c][1]{$t=t+1$}
  \psfrag{d1}[c][c][.9]{$f(d_1^{(j)}) \leq f(d_2^{(j)})$}
  \psfrag{d2}[c][c][.7]{$f(d_2^{(j)}) \leq f(d_\mathrm{sup}^{(j-t)})$}
  \psfrag{d2}[c][c][.68]{\parbox{2.5cm}{\centering  $f(d_\mathrm{sup}^{(j)})$ decreasing over $t_\mathrm{max}$ adapt. steps }}
  \psfrag{d3}[c][c][.68]{\parbox{2.5cm}{\centering  $f(d_\mathrm{sup}^{(j)})$ decreasing over $t_\mathrm{max}$ adapt. steps }}
  \psfrag{d3}[c][c][.8]{$t == t_\mathrm{max}$}
  \psfrag{d5}[c][c][.8]{$t == t_\mathrm{max}$}
  \psfrag{d4}[c][c][.7]{$f(d_1^{(j)}) \leq f(d_\mathrm{sup}^{(j-t)})$}
  \psfrag{f1}[cr][cr]{$f(d_1^{(j)})$}
  \psfrag{f2}[cl][cl]{$f(d_2^{(j)})$}
  \psfrag{y}[cl][cl]{yes}
  \psfrag{n}[cr][cr]{no}
  \includegraphics[scale = .65]{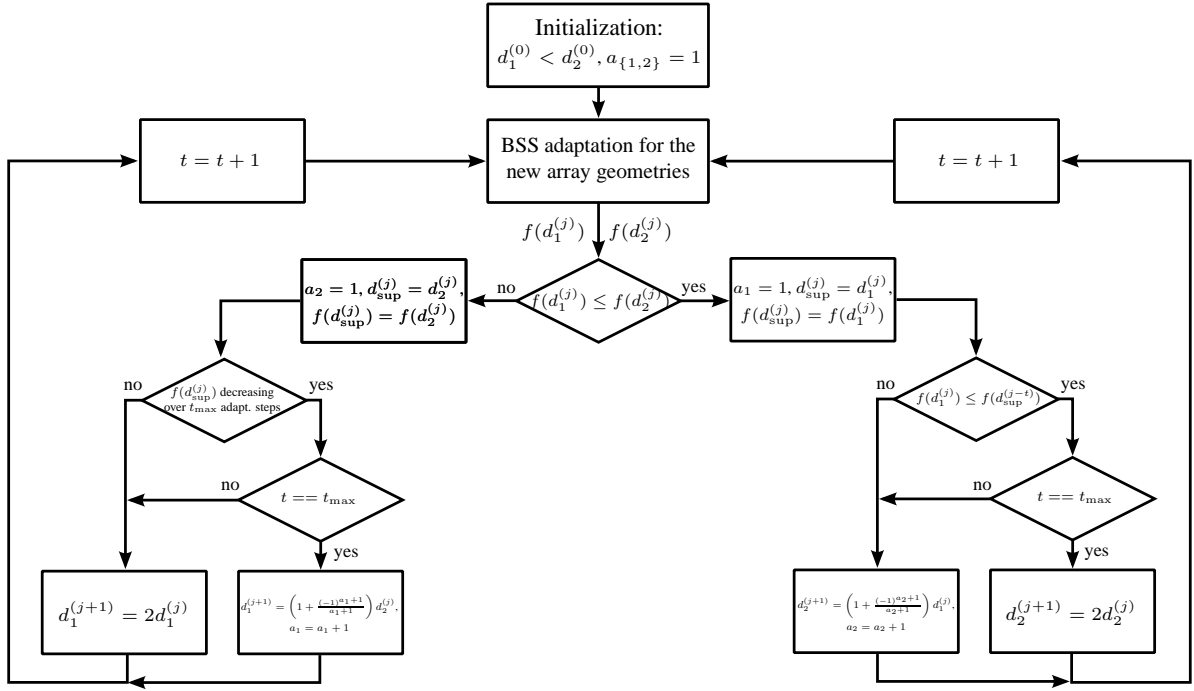}
  \caption{Schematic llustration of the proposed array geometry adaptation algorithm.}
  \label{fig:flowchart_adaptation_strategy}
\end{figure*}

A summary of the proposed adaptation strategy for the array geometry is presented in Fig.~\ref{fig:flowchart_adaptation_strategy} as a flowchart.

During geometry adaptation, the output of the currently superior sub-array is considered as the final output of the signal separation system. 
The decision regarding the superior sub-array is entirely based on the performance measure $f(d_i^{(j)})$. The final output is set to the output of the sub-array that yields the lower value of $f(d_i^{(j)})$.
This comparison is made after each data block that has been processed during the \ac{BSS} adaptation, i.e., after each \ac{BSS} iteration step. 
In order to prevent a toggling between the two \ac{BSS} systems, the final output is only switched from the previously superior sub-array
to the other sub-array, if the performance measure of the latter has been smaller than the performance measure of the previously superior sub-array for at least $m_\mathrm{max}$ data blocks:
$f(d^{(j)}_\mathrm{sup},m) > f(d^{(j)}_\mathrm{inf},m), \, m \in \{1,...,m_\mathrm{max}\}$, where $f(d^{(j)}_{i},m)$ denotes the performance measure obtained from sub-array $i$ after the $m$-th \ac{BSS} iteration step.
For the experiments presented in Section~\ref{sec:experiments}, $m_\mathrm{max}$ was set equal to three.

\section{Experiments}
\label{sec:experiments}
In this section, results from first experiments are presented. To this end, the employed performance measures for \ac{BSS} are introduced in Subsection~\ref{subsec:performanceMeasuresBSS}.
The experimental setup is given in Subsection~\ref{subsec:experimentalSetup} and results are presented in Subsection~\ref{subsec:experimentalResults}.

\subsection{Performance measures for BSS}
\label{subsec:performanceMeasuresBSS}
A very critical aspect of the proposed adaptation algorithm is the need of a performance measure which makes it possible to compare the separation performance of the two sub-arrays blindly and reliably.
The used performance measure is a normalized sum of the weighted \ac{MSC} of the two output signals $y_o, \, o \in \{1,2\}$ of the \ac{BSS} algorithm:
\begin{equation}
	\overline{\mathrm{MSC}} = \frac{1}{\sum_{\nu=0}^{\nu_\mathrm{max}}W(\nu)}\sum \limits_{\nu=0}^{\nu_\mathrm{max}} W(\nu) \frac{|S_{y_{1}y_{2}}(\nu)|^2}{S_{y_{1}y_{1}}(\nu) \cdot S_{y_{2}y_{2}}(\nu)},
	\label{eq:MSC_weighted}
\end{equation}
where $\nu_\mathrm{max}$ denotes the maximum number of frequency bins, $S_{y_{o}y_{o}}(\nu)$ is the auto power spectral density of the two output channels $y_o, \, o \in \{1,2\}$, and 
$S_{y_{o_1}y_{o_2}}(\nu), \, o_1 \neq o_2$ represents the auto power spectral densities of the output channels.
The weighting function $W(\nu)$ at each frequency bin $\nu$ of the \ac{MSC} and is defined as
\begin{equation}
  W(\nu) = \frac{S_{y_{1}y_{1}}(\nu)+S_{y_{2}y_{2}}(\nu)}{2}.
  \label{eq:weightingFunctionMSC}
\end{equation}
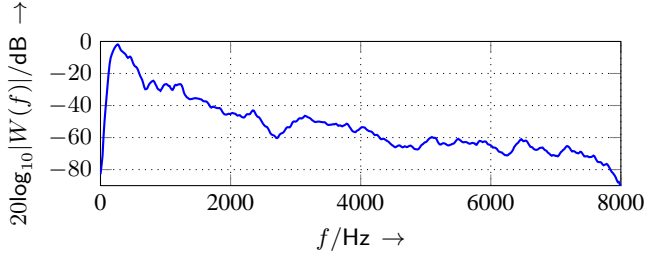
\begin{figure}
  \small
  \begin{tikzpicture}[scale=1]
    \begin{axis}[
      width=8.5cm,height=3.5cm,grid=major,grid style = {dotted,black},
      xlabel={$f/\mathsf{Hz} \, \rightarrow$},
      ylabel={$20\mathsf{log}_{10} |W(f)|/\mathsf{dB} \,\rightarrow$},
      xtick={0,2000,4000,6000,8000},
      xticklabels={0,2000,4000,6000,8000},
      ymin=-90, ymax=0,xmin=0,xmax=8000]
      \addplot[thick,blue,solid] table [x index=0, y index=1]{weightingFunctionMSC.dat};		
    \end{axis}
  \end{tikzpicture}
  \caption{Example of an average speech \ac{PSD} used as window function (\ref{eq:weightingFunctionMSC}) in (\ref{eq:MSC_weighted}).}
  \label{fig:weightingFunctionMSC}
\end{figure}
In Fig.~\ref{fig:weightingFunctionMSC}, an examplary weighting function is illustrated using a logarithmic y-scale. For the sake of simplicity, instead of the frequency bins $\nu$, the corresponding frequency values are given in Fig.~\ref{fig:weightingFunctionMSC}.  
The weighting function accounts for the fact that speech signals have less energy in the higher frequencies and, therefore, the separation performance will be worse in these higher frequency ranges than in lower frequency ranges. 
The $\overline{\mathrm{MSC}}$ is limited to the range of $0 \leq \overline{\mathrm{MSC}} \leq 1$, where $\overline{\mathrm{MSC}}=0$ corresponds to statistically orthogonal signals. 

In addition to the coherence-based measure in (\ref{eq:MSC_weighted}), the well-known \ac{SIR} is used, which requires access to the individual desired and interfering components at the output channels of each \ac{BSS} algorithm. The $\mathrm{SIR}$ is defined as
\begin{equation}
	\mathrm{SIR} = 10\mathrm{log}_{10} \left\{ \frac{|y_{o,\mathrm{d}}[k]|^2}{|y_{o,\mathrm{int}}[k]|^2} \right\} \, \mathsf{dB},
\end{equation}
where $y_{o,\mathrm{d}}[k]$ and $y_{o,\mathrm{int}}[k]$ denote the desired and interfering signal components at time instant $k$ at the $o$-th output channel. Here, the signal that is suppressed in one output channel is considered to be the interfering signal in this output channel. Thus, for each output channel, high \ac{SIR} levels are desirable.

\subsection{Experimental setup}
\label{subsec:experimentalSetup}
\begin{figure}[htb]
  \hspace{5mm}
  \small
  \psfrag{d1}[cr][cr][.85]{$d_1^{(j)}$}
  \psfrag{d2}[cr][cr][.85]{$d_2^{(j)}$}
  \psfrag{d3}[cr][cr][.85]{$20^\circ$}
  \psfrag{d4}[cl][cl][.85]{$1.0\, \mathsf{m}$}
  \psfrag{d5}[cr][cr][.85]{$-20^\circ$}
  \psfrag{d6}[c][c][.85]{\parbox{1.5cm}{\centering \ac{BSS}\\ sub-array~1}}
  \psfrag{d7}[c][c][.85]{\parbox{1.5cm}{\centering \ac{BSS}\\ sub-array~2}}
  \psfrag{d8}[c][c][.85]{Decision}
  \psfrag{d9}[cl][cl][.85]{$\overline{\mathrm{MSC}}_1$, $\mathrm{SIR_{mean,1}}$}
  \psfrag{d10}[cl][cl][.85]{$\overline{\mathrm{MSC}}_2$, $\mathrm{SIR_{mean,2}}$}
  \psfrag{o1}[cl][cl][.85]{$\mathrm{SIR_{mean,out}}$}
  \includegraphics[scale=.8]{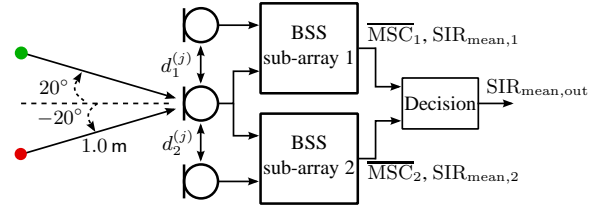}
  \caption{Illustration of the tested multi-speaker scenario and employed measures of performance.}
  \label{fig:scenario}
\end{figure}

The adaptation algorithm has been tested in a two-source environment with a male and a female speaker of equal power
located at $20^\circ$ and $-20^\circ$ at a distance of $1.0 \, \mathsf{m}$ from the microphone array, as illustrated in Fig.~\ref{fig:scenario}.
The \acp{RIR}, modeling the propagation from the sources to the microphones, were simulated using the image method proposed by Allen and Berkley \cite{Allen:1979fk}.
The simulated room was of dimensions $( 4.5 \, \mathsf{m} \times 4.5 \, \mathsf{m} \times 2.5 \, \mathsf{m} )$ with a reverberation time of $T_{60} \approx 200\, \mathsf{ms}$, corresponding to a critical distance \cite{Lerch:2009fk} of apporximately $0.9 \, \mathsf{m}$.
The microphone signals were synthesized by convolving clean speech signals of sampling rate $f_\mathrm{s}=16 \, \mathsf{kHz}$ with the simulated \acp{RIR}. The array geometry adaptation was simulated for continuous-speech signals of length $30$ seconds.
%
For the \ac{BSS} algorithms a filter length of $L=1024$ was used and the power-spectral densities required for the calculation of $\overline{\mathrm{MSC}}$~(\ref{eq:MSC_weighted}) were estimated using the Welch Method with a window length of $4L$ samples and $50\%$ overlap.
Across the blocks, a recursive averaging has been applied for the estimated power-spectral densities.

\subsection{Experimental results}
\label{subsec:experimentalResults}
\begin{figure}[h] 
  \scriptsize
  \centering
  \subfigure[$\mathrm{SIR_{mean}}$ obtained from sub-array~2.]{
    \begin{tikzpicture}[scale=1]
      \begin{axis}[
	width=4.5cm,height=3.5cm,grid=major,grid style = {dotted,black},
	xlabel={Microphone spacings $d_2/\mathsf{m}$},
	ylabel={$\mathrm{SIR_{mean}}/\mathsf{dB} \,\rightarrow$},
	xtick={1,2,3,4},
	xticklabels={0.10,0.15,0.20,0.25},
	ymin=3, ymax=8.75,xmin=1,xmax=4]
	\addplot[thick,blue,solid,mark=x] table [x index=0, y index=1]{comparison_SIRmean_MSC_200ms_L1024.dat};
      \end{axis}
    \end{tikzpicture}
    \label{fig:finalMSC_SIRmean_a}
  }
  \hfill
  \subfigure[$\overline{\mathrm{MSC}}$ obtained from sub-array~2.]{
    \begin{tikzpicture}[scale=1]
      \begin{axis}[
	width=4.5cm,height=3.5cm,grid=major,grid style = {dotted,black},
	xlabel={Microphone spacings $d_2/\mathsf{m} \, \rightarrow$},
	ylabel={$\overline{\mathrm{MSC}} \,\rightarrow$},
	xtick={1,2,3,4},
	xticklabels={0.10,0.15,0.20,0.25},
	ymin=0.075, ymax=.225,xmin=1,xmax=4]
	\addplot[thick,blue,solid,mark=x] table [x index=0, y index=2]{comparison_SIRmean_MSC_200ms_L1024.dat};		
      \end{axis}
    \end{tikzpicture}
  \label{fig:finalMSC_SIRmean_b}
  }
  \caption{Comparison of $\mathrm{SIR_{mean}}$ and $\overline{\mathrm{MSC}}$ obtained from sub-array~2 for different microphone spacings.}
  \label{fig:finalMSC_SIRmean}
\end{figure}
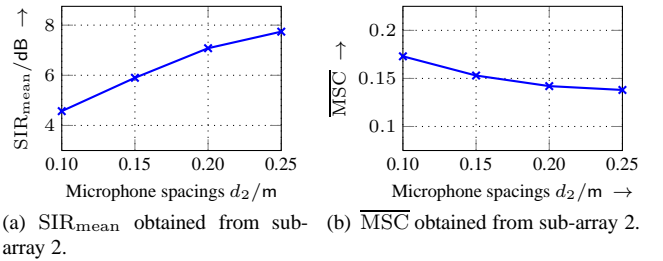

In Fig.~\ref{fig:finalMSC_SIRmean}, a comparison between $\mathrm{SIR_{mean}}$ and $\overline{\mathrm{MSC}}$ obtained from sub-array~2 for different microphone spacings is given. 
Here, $\mathrm{SIR_{mean}}$ denotes the arithmetic average of the output \acp{SIR} at the two output channels of the \ac{BSS} algorithm. As can be seen, higher $\mathrm{SIR_{mean}}$ levels correspond to lower $\overline{\mathrm{MSC}}$ values and vice versa.
Thus, the $\overline{\mathrm{MSC}}$ is appears to be well suited to evaluate the separation performance of the \ac{BSS} algorithm blindly and reliably.

\begin{figure}[h] 
  \small
  \subfigure[$\overline{\mathrm{MSC}}$ obtained from both sub-arrays during \ac{BSS} and geometry adaptation.]{
    \scriptsize
    \begin{tikzpicture}[scale=1]
      \def\lx{.05} 
      \def\ly{1.98} 
      \begin{axis}[
	width=8.5cm,height=4cm,grid=major,grid style = {dotted,black},
	xlabel={Time $t/\mathsf{s} \, \rightarrow$},
	ylabel={$\overline{\mathrm{MSC}} \,\rightarrow$},
	xtick={1,19,38,57},
	xticklabels={0,10,20,30},
	ymin=0,ymax=1.3,xmin=1,xmax=57]
	\addplot[thick,blue,dashed] table [x index=0, y index=1]{results_adaptation_200ms_L1024.dat};
	\addplot[thick,red,solid] table [x index=0, y index=2]{results_adaptation_200ms_L1024.dat};
      \end{axis}
      \draw[fill=white] (\lx,\ly) rectangle (2.8+\lx,0.4+\ly);
      \draw[red,thick,solid] (.2+\lx,0.15+\ly) -- +(0.3,0) node[anchor=mid west,black] {\scriptsize $\overline{\mathrm{MSC}}_1$};
      \draw[blue,thick,dashed] (1.5+\lx,0.15+\ly) -- +(0.3,0) node[anchor=mid west,black] {\scriptsize $\overline{\mathrm{MSC}}_2$};
    \end{tikzpicture}
    \label{fig:adaptation_MSC_SIRmean_d_a}
  }
  \vspace{1mm}
  \subfigure[$\mathrm{SIR_{mean}}$ obtained from both sub-arrays during \ac{BSS} and geometry adaptation.]{
    \scriptsize
  \hspace{.08mm}
    \begin{tikzpicture}[scale=1]
      \def\lx{.05} 
      \def\ly{1.98} 
      \begin{axis}[
	width=8.5cm,height=4cm,grid=major,grid style = {dotted,black},   
	xlabel={Time $t/\mathsf{s} \, \rightarrow$},
	ylabel={$\mathrm{SIR_{mean}}/\mathsf{dB} \,\rightarrow$},
	xtick={1,19,38,57},
	xticklabels={0,10,20,30},
	ymin=-1,ymax=14,xmin=1,xmax=57]
	ytick={0,5,10,15,20}
	\addplot[thick,red,solid] table [x index=0, y index=4]{results_adaptation_200ms_L1024.dat};
	\addplot[thick,green,solid] table [x index=0, y index=5]{results_adaptation_200ms_L1024.dat};
	\addplot[only marks,mark=x,color=green,thick] table [x index=0, y index=5]{results_adaptation_200ms_L1024_markers.dat}; 
	\addplot[thick,blue,dashed] table [x index=0, y index=3]{results_adaptation_200ms_L1024.dat};
      \end{axis}
      \draw[fill=white] (\lx,\ly) rectangle (6+\lx,0.4+\ly);
      \draw[red,thick,solid] (0.2+\lx,0.2+\ly) -- +(0.3,0) node[anchor=mid west,black] {\scriptsize $\mathrm{SIR_{mean,1}}$};
      \draw[blue,thick,dashed] (2.1+\lx,0.2+\ly) -- +(0.3,0) node[anchor=mid west,black] {\scriptsize $\mathrm{SIR_{mean,2}}$};
      \draw[thick,green,solid] (4+\lx,0.2+\ly) -- +(0.3,0) node[anchor=mid west,black] {\scriptsize $\mathrm{SIR_{mean,\mathrm{out}}}$};
      \draw[thick,green,solid] (4.15+\lx,0.2+\ly) node{x};
    \end{tikzpicture}
    \label{fig:adaptation_MSC_SIRmean_d_b}
  }
  \vspace{1mm}
  \subfigure[Microphone spacings $d_i^{(j)}$ after each geometry iteration step $j$.]{
    \scriptsize
  \hspace{-3.5mm}
    \begin{tikzpicture}[scale=1]
      \def\lx{4} 
      \def\ly{.25} 
      \begin{axis}[
	width=8.5cm,height=4cm,grid=major,grid style = {dotted,black},
	xlabel={Iteration step $j \, \rightarrow$},
	ylabel={$d_i^{(j)}/\mathsf{m} \,\rightarrow$},
	xtick={1,19,38,57},
	xticklabels={0,1,2,3},
	ytick={1,2,3,4},
	yticklabels={0.15,0.20,0.30,0.45},
	ymin=.5,ymax=4.5,xmin=1,xmax=57]
	\addplot[thick,blue,dashed] table [x index=0, y index=6]{results_adaptation_200ms_L1024.dat};
	\addplot[thick,red,solid] table [x index=0, y index=7]{results_adaptation_200ms_L1024.dat};
      \end{axis}
      \draw[fill=white] (\lx,\ly) rectangle (2.5+\lx,0.4+\ly);
      \draw[red,thick,solid] (0.2+\lx,0.15+\ly) -- +(0.3,0) node[anchor=mid west,black] {\scriptsize $d_1^{(j)}$};
      \draw[blue,thick,dashed] (1.5+\lx,0.15+\ly) -- +(0.3,0) node[anchor=mid west,black] {$d_2^{(j)}$};
    \end{tikzpicture}
    \label{fig:adaptation_MSC_SIRmean_d_c}
  }
  \caption{Illustration of the adaptation of the microphone spacings $d_i^{(j)}$.}
  \label{fig:adaptation_MSC_SIRmean_d}
\end{figure}
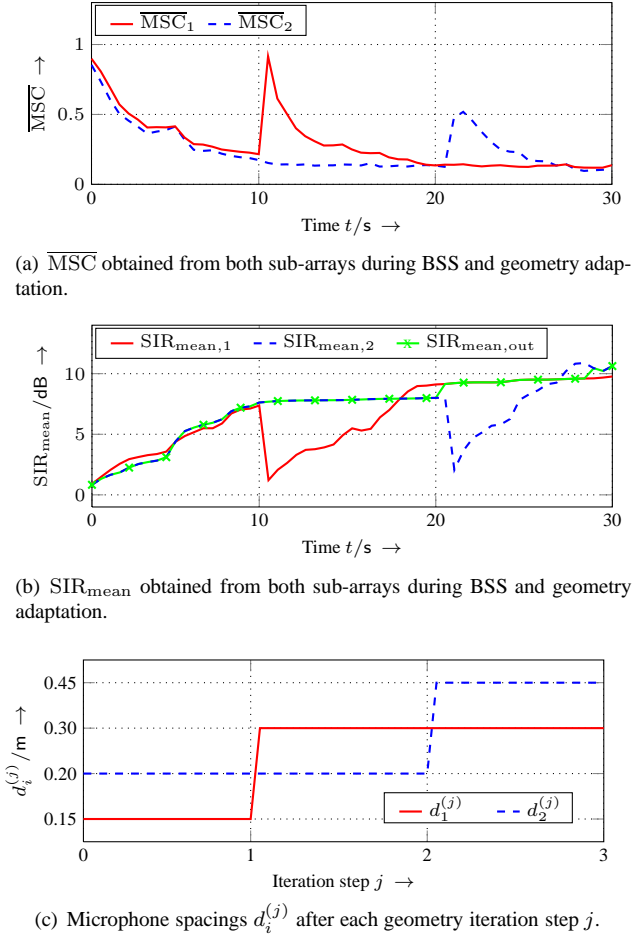

In Fig.~\ref{fig:adaptation_MSC_SIRmean_d}, the adaptation progress of the two microphone spacings $d_i^{(j)}, \, i\in\{1,2\}$ is illustrated. 
For the sake of simplicity, all employed performance measures are also indicated in Fig.~\ref{fig:scenario}.
In Fig.~\ref{fig:adaptation_MSC_SIRmean_d_a} and \ref{fig:adaptation_MSC_SIRmean_d_b}, the adaptation of the \ac{BSS} algorithm to the current microphone positions of each sub-array is depicted over time. 
The obtained $\overline{\mathrm{MSC}}$ values of each sub-array are given in Fig.~\ref{fig:adaptation_MSC_SIRmean_d_a}, where $\overline{\mathrm{MSC}_i}$ denotes the value obtained from the $i$-th sub-array.
After segments of length of five seconds, the performance measures are compared and the microphone spacings are adapted as described above.
In Fig.~\ref{fig:adaptation_MSC_SIRmean_d_b}, the corresponding $\mathrm{SIR_{mean}}$ levels are given. $\mathrm{SIR_{mean,out}}$, which is represented by the solid green curve with x-markers, denotes the $\mathrm{SIR_{mean}}$ 
level at the output of the extraction algorithm. It is equal to the $\mathrm{SIR_{mean}}$ level that is obtained by the superior sub-array.
The decision regarding the superior sub-array is entirely based on $\overline{\mathrm{MSC}}_1$ and $\overline{\mathrm{MSC}}_2$ and is described in Section~\ref{sec:proposedAdaptationAlgorithm}.
The employed microphone spacings are illustrated in Fig.~\ref{fig:adaptation_MSC_SIRmean_d_c}, where the horizontal axis represents the current iteration step $j$ of the array geometry adaptation.
The algorithm starts at $d_1^{(0)}=0.15\,\mathsf{m}$ and $d_2^{(0)}=0.20\,\mathsf{m}$, respectively. After the first iteration step, $d_1$ is adapted according to (\ref{eq:adaptationStrategy}), since sub-array~$2$ yields
a better separation performance than sub-array~1, as can be seen from Fig.~\ref{fig:adaptation_MSC_SIRmean_d_a}. 
At $j=2$, $\overline{\mathrm{MSC}}_1$ is smaller than $\overline{\mathrm{MSC}}_2$, thus $d_2$ is adapted and $d_1$ remaines unchanged.
Looking at $\mathrm{SIR_{mean,out}}$ in Fig.~\ref{fig:adaptation_MSC_SIRmean_d_b}, it can be observed that the mean \ac{SIR} at the output is steadily increasing over iteration steps $j$, corresponding to an improved performance of the signal separation due to the adapted microphone array geometry.

\section{Summary and Outlook}
\label{sec:summary_outlook}
In this paper, the generic concept of an adaptive microphone array geometry was introduced for application with humanoid robots. It was demonstrated that it is an efficient method to increase the performance of BSS.
By using this adaptation algorithm in the context of the \ac{BSE} scheme \cite{lnt2012-62}, it is expected to provide better noise estimates and thus increase the signal extraction performance. 
Consequently, \ac{ASR} scores are expected to improve significantly in complex and adverse acoustic scenarios, leading to a significant reduction of task completion time in human/robot interaction.
Future work will include a more detailed investigation and possible refinements of the proposed performance measure, robust optimization of the current adaptation mechanism with recorded microphone data in different acoustic environments, 
investigation of alternative adaptation approaches, as well as the generalization to a larger number of movable microphones.

\vfill\pagebreak

\label{sec:refs}

\bibliographystyle{IEEEbib}
\bibliography{bibliography}

\end{document}